# Stock prices, inflation and inflation uncertainty in the U.S.: Testing the long-run relationship considering Dow Jones sector indexes

ClaudiuTiberiuALBULESCU[a*], Christian AUBIN[b], Daniel GOYEAU[b]

[a] Management Department, Politehnica University of Timisoara

[b] CRIEF, University of Poitiers

**Abstract**

We test for the long-run relationship between stock prices, inflation and its uncertainty for different U.S. sector stock indexes, over the period 2002M7 to 2015M10. For this purpose we use a cointegration analysis with one structural break to capture the crisis effect, and we assess the inflation uncertainty based on a time-varying unobserved component model. In line with recent empirical studies we discover that in the long-run, the inflation and its uncertainty negatively impact the stock prices, opposed to the well-known Fisher effect. In addition we show that for several sector stock indexes the negative effect of inflation and its uncertainty vanishes after the crisis setup. However, in the short-run the results provide evidence in the favor of a negative impact of uncertainty, while the inflation has no significant influence on stock prices, except for the consumption indexes. The consideration of business cycle effects confirms our findings, which proves that the results are robust, both for the long- and the short-run relationships.

[*] Corresponding author. E-mail addresses: claudiu.albulescu@upt.ro, claudiual@yahoo.com.

# 1. Introduction

Studies regarding the impact of inflation on asset prices have a long tradition in the financial field. The starting point is considered the "The Theory of Interest" by Fisher (1930), who advanced the idea that expected nominal return of an asset should equal the expected real return and the expected rate of inflation (the so-called "Fisher effect"). However, the empirical literature hardly succeeded to demonstrate this effect, even if this was the common view before the 1970s. As such, noteworthy works demonstrated on contrary, that the inflation negatively impact the stock returns (Bodie, 1976; Jaffe and Mandelker, 1976; Nelson, 1976; Fama and Schwert, 1977), while subsequent theoretical and empirical papers found explanations for a potential negative relationship.

First, Modigliani and Cohn (1979) suggest that stock market investors fail to understand the inflation's effects on the nominal cash flow, and during the periods of high inflation, the valuation errors induce an undervaluation in stocks, and *vice-versa*. Second, Feldstein (1980) shows that the inflation generates artificialcapital gains which are subject to taxation.Therefore, in the presence of tax system distortions, if the inflation increases, the firms face higher tax liabilities. In this context, rational investors reduce common stock valuation to takeinto account the effect of inflation, which negatively affect the stock prices (tax-effects hypothesis).Third, Fama (1981) shows that the inverse relationship between real stock returns and expected inflation is generated by a positive relationship between equity market and the output growth, combined with a negative relationship between expected inflation and real economic activity (the proxy hypothesis). Fourth, Geske and Roll (1983) and Kaul (1987) explain the negative relationship using the argument of a counter-cyclical monetary policy. They suggest that positive shocks to real output generate monetary tightening which reduces the inflation, while triggering anincrease in stock prices. Fifth, Hess and Lee (1999) explain the negative relationship as a combination of demand and supply disturbances.Sixth, Boudoukh and Richardson (1993) followed by others, provide a more complex explanation, arguing that stock returns are inversely related to realized and expected inflation in the short-run, but may be positively related to inflation in the long-run. However, using a similar framework, Sharpe (2002) draws a new perspective on the relationship between stock prices and inflation, underlining the existence of a potential negative relationship in the long-run also. He shows that a rise in expected inflation is accompanied by either a decline in expected long-run real earnings, or by a rise in the long-run real return, or both.

A different strand of literature put accent on the role of inflation uncertainty to explain the link between stock prices and inflation. Starting with Levi and Makin (1979) and Kolluri (1982), economistsdescribe the Friedman effect[1] and its role in a Fisherian-type relationship.The Friedman's second reasoning revealed in his Nobel lecture shows that the inflation uncertainty negatively impact the real output. In this context, Hu and Willett (2000) and Park and Ratti (2000) demonstrate thatexpectations of an output decline depress current stock prices, effect reinforced by greater economic uncertainty during high inflationary periods. Analternative explanation about the role of inflation uncertainty is given by Amer (1994). According to Amer (1994), a rise in inflation increase riskiness of assets and therefore the expected rate of return, as an indirect effect of an increase uncertainty. *Ceteris paribus*, the stock prices drop in present if an increase in future returns is expected (Tobin, 1958; Campbell, 1991).[2]

However, the existing empirical literature addressing the relationship between inflation and stock prices makes abstraction of inflation uncertainty, with few exceptions (for a discussion see Azar, 2013). Moreover, the mixed results and the lack of distinction between the short- and long-run horizons in estimating the nexus between stock prices, inflation and its uncertainty require supplementary investigations. Therefore, the present paper's contributions to the empirical literature are three fold.

First, we focus on the characteristics of the inflation and its uncertainty, which prove to be non-stationary processes, and on the endogeneity which exists between stock prices, inflation and its uncertainty.Given these evidence, we use a cointegration approach, focusing on the recent period and employing U.S. statistics over the period 2002M7 to 2015M10. In order to underline the effect of the recent financial crisis we resort to the Gregory-Hansen cointegration test with one structural break (Gregory and Hansen, 1996a). The cointegration analysis allows to investigate the above-mentioned relationship at different time-horizons, and to explore the hypotheses advanced by Sharpe (2002). We focus on stock price level and not on returns as Cochran and Defina (1993) and Alexakis et al. (1996) did. Indeed

---

[1] The link between inflation and its uncertainty on the one hand, and between the inflation uncertainty and output on the other hand, became famous with the Friedman's Nobel lecture (Friedman, 1977). The first hypothesis of Friedman, showing the role of inflation uncertainty in explaining the level of inflation, was formalized by Ball (1992) (we call this Friedman – Ball hypothesis). Afterwards, several competing hypothesis where advanced and become famous, showing the positive impact of inflation on its uncertainty (Cukierman and Meltzer, 1986), or on contrary, a negative relationship where the inflation leads uncertainty (Pourgerami and Maskus, 1987), or where the inflation is leaded by its uncertainty (Holland, 1995).
[2] Campbell (1991) explains this reasoning by the fact that, if the stock returns are expected to rise in the distant future and if the path of dividends is fixed, then the stock price must drop in present to allow a rise in the future.

Amer(1994) shows that a modification of expected returns (which are not observable), leads to a modification of the present stock prices.

Second, different from previous studies which associate the inflation uncertainty with the inflation variability and use different generalized autoregressive conditional heteroskedasticity (GARCH)-type models, we consider that expected inflation is generally unobservable (see Cukierman and Meltzer, 1986; Kolluri and Wahab, 2008). Further, the use of GARCH-type models to measure inflation uncertainty requires stationary time-series, which is not the case for the inflation rate. Therefore, we use an alternative measure of inflation uncertainty based on theunobserved component model with stochastic volatilityproposed by Stock and Watson's (2007). In this model the inflation is decomposed in a trend component and the inflation gap, which is associated with the uncertainty.

Third, existing studies, however, do not focus on stock markets sector indexes. Nevertheless, the impact of inflation and its uncertainty on stock prices might differ for various economic sectors (i.e. free-market established prices versus regulated prices, raw material versus final consumption goods, etc.). Therefore, we use in our analysis ten Dow Jones (DJ) sector indexes to assess the potentially different effect of inflation and its uncertainty.Finally, in order to see if the results are influenced or not by the phases of business cycle, we perform a robustness check analysis including in our estimations the economic growth rate.

The rest of the paper is structured as follows. Section 2 presents a brief literature review of papers addressing the link between stock prices and inflation on the one hand, and between stock prices, inflation and its uncertainty on the other hand. Section 3 describes the research methodology and data. Section 4 presents the empirical results. Section 5 realizes a robustness check while the last section concludes.

## 2. Empirical literature review

The empirical evidence suggests the presence ofcomplexities, regarding the stock prices – inflation nexus, and it is heavily oriented toward the U.S. case. While oldest empirical studies examined this relationship at relatively short horizons, subsequent works focused on longer time-horizons (Boudoukh and Richardson, 1993; Solnik and Solnik, 1997; Schotman and Schweitzer, 2000), and found evidence for the Fisher effect. On contrary, other studies like Engsted and Tanggaard (2002) discovered that the Fisher effect diminishes with the time horizon increase.

Recent empirical works focus on nonlinearities which may exist between stock prices, inflation and its uncertainty(Boyd, 2001; Kim, 2003; Liu et al., 2005;Maghyereh, 2006; Karagianni and Kyrtsou, 2011). While for example Liu et al. (2005) apply a regime-switching model, Maghyereh (2006) resorts to a nonparametric cointegration test. Further, Karagianni andKyrtsou (2011) use a recurrence quantification analysis and a series of tests for structural breaks and nonlinear causality, documenting negative nonlinear linkages between the inherent dynamics of inflation and stock returns. Further, several studies combines the time and frequency domains, resorting to wavelets. Kim and In (2005) use a six level decomposition of the U.S. inflation and stock returns and show that there is a positive relationship at the shortest and longest time-scales, while at intermediate scales stock return and inflation are negatively correlated. More recently, Tiwari et al. (2015) employ a continuous wavelet transform and document a positive relation for higher time scales, which however lacks in robustness when a different measure of inflation is used.

Few studies investigate the link between stock prices, inflation and its uncertainty (see Alexakis et al., 1996; Azar, 2013). Along these, Alexakis et al. (1996) examines the link between inflation uncertainty and stock prices for a group of developed and emerging economies over the period 1980M1 to 1993M12, and report a negative relationship. Recently, Azar (2013) shows thatinflation uncertainty dominates the inflation in explaining stock prices in the case of the U.S. over the time-span 1950M1-2011M3. However, both variables become redundant when other fundamental variables are included in the regression, which shows that neither the inflation, nor its uncertainty, have a strong impact on stock prices.

In line with these papers we investigate the nexus between stock prices, inflation and its uncertainty for the U.S. However, different from previous works, we focus on the long-run relationship and perform a sector-level analysis. Moreover, we use a newly proposed measure of inflation uncertainty, based on a time-varying unobserved component model.

### 3. Methodology and data

*3.1. Measuring the inflation uncertainty*

While many researches associate the inflation uncertainty with the inflation volatility and use different GRACH-type approaches for computing the volatility, other studiesrelate the inflation uncertainty to the inflation gap (Stock and Watson, 2007; Cogley et al., 2010). In the second case the unobserved component (UC) model is employed.

Stock and Watson (2007) propose a generalized form of the UC model in which the variances of the permanent and transitory disturbances evolve randomly over time. The new model, called the unobserved component model with stochastic volatility (UC-SV), is the following:

$$\pi_t = \tau_t + \eta_t, \text{ where } \eta_t = \sigma_{\eta,t}\zeta_{\eta,t} \tag{1}$$

$$\tau_t = \tau_{t-1} + \varepsilon_t, \text{ where } \varepsilon_t = \sigma_{\varepsilon,t}\zeta_{\varepsilon,t} \tag{2}$$

$$ln\sigma_{\eta,t}^2 = ln\sigma_{\eta,t-1}^2 + \upsilon_{\eta,t} \tag{3}$$

$$ln\sigma_{\varepsilon,t}^2 = ln\sigma_{\varepsilon,t-1}^2 + \upsilon_{\varepsilon,t} \tag{4}$$

where: $\pi_t$ is the level of inflation, $\tau_t$ is the inflation stochastic trend, and the $\eta_t$ represents the serially uncorrelated disturbance (inflation gap), with the property $\zeta_t = (\zeta_{\eta,t}, \zeta_{\varepsilon,t})$ is i.i.d. $N(0, I_2)$, $\upsilon_t = \upsilon_{\eta,t}, \upsilon_{\varepsilon,t}$ is i.i.d. $N(0, \gamma I_2)$, $\zeta_t$ and $\upsilon_t$ are independently distributed and $\gamma$ is a scalar parameter which controls the smoothness of the stochastic volatility process set by Stock and Watson (2007) at 0.04.

### *3.2. Gregory-Hansen cointegration test*

Our general equation is:

$$DJ = c + \alpha_1 I + \alpha_2 U + \varepsilon_t \tag{5}$$

where: $DJ$ is the natural log of the DJ sector indexes; $c$ is the intercept; $I$ is the CPI inflation; $U$ represents the inflation uncertainty; $\varepsilon_t$ are the error terms.

In order to capture the long-run relationship between variables, usually $c$ and $\alpha$ are considered time-invariant. However, Gregory and Hansen (2006a) consider that, if the cointegration holds over some periods of time, it may shift toward a new long-run relationship. They treat the timing and shifts as unknown and allows changes in the intercept and slope, defining the following dummy variable:

$$\varphi_{t\tau} = \begin{cases} 0, & if \ t \leq [n\tau] \\ 1, & if \ t > [n\tau] \end{cases} \tag{6}$$

where: $\tau \in (0,1)$ is an unknown parameter denoting the timing of the change point.

The cointegration test proposed by Gregory and Hansen (2006a) accommodates a single endogenous break. In order to identify the timing of the change point, a cointegration test is computed for each possible shift, and the smallest value is retained across all possible break points.

Gregory and Hansen (2006a) propose three models with assumptions about structural breaks in the intercept and slope. Afterwards, in Gregory and Hansen (2006b) a forth model is advanced, allowing for breaks in the trend also. Based on these four models, our equations become:

Model 1 – Gregoryand Hansen's (1996a) test: Level shift (GH-LS)

$$DJ = c_1 + c_2\varphi_{t\tau} + \alpha_1 I + \alpha_2 U + \varepsilon_t \qquad (7)$$

Model 2 – Gregory and Hansen's (1996a) test: Level shift with trend (GH-LST)

$$DJ = c_1 + c_2\varphi_{t\tau} + \beta t + \alpha_1 I + \alpha_2 U + \varepsilon_t \qquad (8)$$

Model 3 – Gregory and Hansen's (1996a) test: Regime shift (GH-RS)

$$DJ = c_1 + c_2\varphi_{t\tau} + \alpha_1 I + \alpha_2 I\varphi_{t\tau} + \alpha_3 U + \alpha_4 U\varphi_{t\tau} + \varepsilon_t \qquad (9)$$

Model 4 – Gregory and Hansen's (1996b) test: Regime shift with trend change (GH-RST)

$$DJ = c_1 + c_2\varphi_{t\tau} + \beta_1 t + \beta_1 t\varphi_{t\tau} + \alpha_1 I + \alpha_2 I\varphi_{t\tau} + \alpha_3 U + \alpha_4 U\varphi_{t\tau} + \varepsilon_t \qquad (10)$$

The test statistics computation in Gregory and Hansen (1996a) is based on the Ordinary Least Square (OLS) approach and each model yields the residuals $\hat{\varepsilon}_{t\tau}$, where the subscript $\tau$ shows that the residual sequence depends on the choice of the change point $\tau$. The first-order serial correlation coefficient is:

$$\hat{\rho}_\tau = \sum_{t=1}^{n-1} \hat{\varepsilon}_{t\tau}\, \hat{\varepsilon}_{t+1\tau} / \sum_{t=1}^{n-1} \hat{\varepsilon}_{t\tau}^2 \qquad (11)$$

The second-stage residuals are defined as $\hat{v}_{t\tau} = \hat{\varepsilon}_{t\tau} - \hat{\rho}_\tau \hat{\varepsilon}_{t-1\tau}$, while the estimate of long-run variance of $\hat{v}_{t\tau}$ is $\hat{\sigma}_\tau^2 = \hat{\gamma}_\tau(0) + 2\hat{\lambda}_\tau$. The estimate of the bias-corrected first-order serial correlation coefficient is:

$$\hat{\rho}_t^* = \sum_{t=1}^{n-1}(\hat{\varepsilon}_{t\tau}\, \hat{\varepsilon}_{t+1\tau} - \hat{\lambda}_\tau) / \sum_{t=1}^{n-1} \hat{\varepsilon}_{t\tau}^2 \qquad (12)$$

Gregory and Hansen (1996a) propose three test statistics, namely two Phillips statistics and one Augmented Dickey–Fuller (ADF statistic). The Phillips statistics are the following:

$$Z_\alpha(\tau) = n(\hat{\rho}_t^* - 1) \qquad (13)$$

$$Z_t(\tau) = (\hat{\rho}_t^* - 1)/\hat{s}_\tau \qquad (14)$$

where: $\hat{s}_\tau^2 = \hat{\sigma}_\tau^2 / \sum_1^{n-1} \hat{\varepsilon}_{t\tau}^2$.

The ADF statistics is calculated by regressing $\Delta\hat{\varepsilon}_{t\tau}$ upon $\hat{\varepsilon}_{t-1\tau}$ and $\Delta\hat{\varepsilon}_{t-1\tau}, \ldots, \Delta\hat{\varepsilon}_{t-K\tau}$ for suitably chosen lag truncation $K$. As such, the ADF statistics is the *t*-statistics for the regressor $\hat{\varepsilon}_{t-1\tau}$, denoted by:

$$ADF(\tau) = tstat(\hat{\varepsilon}_{t-1\tau}) \qquad (15)$$

*3.3. Data*

The CPI inflation and industrial production growth rate are extracted from FED St. Louis (FRED database).The DJ sector indexes data are obtained from Investing.com database and are available starting with 2002M7, on a monthly frequency. Therefore our sample covers the period 2002M7 to 2015M10. The tenDJ sector indexes, classified according to the proprietary classification systems'industries are: DJ basic materials index (DJUSBM), DJ consumer goods index (DJUSNC), DJ consumer services index (DJUSCY), DJ financial index (DJUSFN), DJ health care index (DJUSHC), DJ industrials index (DJUSIN), DJ oil and gas index (DJUSEN), DJ technology index (DJUSTC), DJ telecommunication index (DJUSTL) and DJ utilities index (DJUSUT).

In order to check if there is a long-run relationship between the stock prices, the inflation and its uncertainty, our series shall be I(1). For testing thus the presence of unit roots, we resort to the classic ADF and Phillips–Perron (PP) tests. Table 1 shows that the ADF and PP tests cannot reject the null of unit root presence in level, for any of the selected variables. On contrary the stationarity is documented in the first difference for all variables, although to a smaller extent for the inflation uncertainty. We conclude then that our variables are non-stationary in level but stationary in first difference and we proceed to the cointegration analysis.

Table 1. Unit root tests

| Variables | ADF test | | PP test | |
| --- | --- | --- | --- | --- |
| | *Level* | *First difference* | *Level* | *First difference* |
| DJUSBM | -1.85 | -10.6*** | -2.10 | -10.7*** |
| DJUSNC | -0.08 | -11.5*** | -0.19 | -11.5*** |
| DJUSCY | 0.19 | -11.5*** | 0.07 | -11.5*** |
| DJUSFN | -1.21 | -10.2*** | -1.50 | -10.2*** |
| DJUSHC | 0.57 | -11.5*** | 0.56 | -11.5*** |
| DJUSIN | -0.90 | -10.7*** | -1.09 | -10.7*** |
| DJUSEN | -2.10 | -12.4*** | -2.09 | -12.4*** |
| DJUSTC | -1.04 | -12.0*** | -1.05 | -11.9*** |
| DJUSTL | -2.17 | -12.5*** | -2.04 | -12.6*** |
| DJUSUT | -1.42 | -11.5*** | -1.45 | -11.5*** |
| I | -2.03 | -8.07*** | -2.45 | -7.85*** |
| U | -1.00 | -2.73* | -1.97 | -3.47** |
| IP | -1.49 | -9.95*** | -2.16 | -10.4*** |

*Notes: (i) \*, \*\* and \*\*\* implies significance at 10%, 5% and 1% levels respectively and rejecting H0 of unit root presence; (ii) the natural log of stock indexes is considered.*

## 4. Empirical findings

*4.1. Cointegration tests*

The results of the Gregory-Hansen cointegration tests are presented in Table 2. We consider that a long-run relationship exists if two out of three tests confirm the cointegration at list at 10% level of significance.

Table 2. Gregory-Hansen cointegration tests for stock prices, inflation and uncertainty

| Models | ADF | Break | $Z_t$ | $Z_a$ | Break |
|---|---|---|---|---|---|
| **DJUSBM** | | | | | |
| GH-LS | -4.38 | 2006M4 | -4.17 | -25.5 | 2006M5 |
| GH-LST | -5.03 | 2010M12 | -4.86 | -35.9 | 2011M3 |
| GH-RS | -5.05 | 2007M6 | -5.26* | -43.2 | 2008M7 |
| GH-RST | -6.39** | 2009M12 | -6.18** | -56.1 | 2009M4 |
| **DJUSNC** | | | | | |
| GH-LS | -3.39 | 2005M2 | -3.91 | -24.1 | 2004M6 |
| GH-LST | -6.09*** | 2009M12 | -6.19*** | -56.0** | 2010M1 |
| GH-RS | -3.90 | 2005M2 | -3.78 | -25.9 | 2008M8 |
| GH-RST | -6.06** | 2009M12 | -6.12** | -56.0 | 2009M4 |
| **DJUSCY** | | | | | |
| GH-LS | -3.20 | 2013M6 | -3.89 | -23.8 | 2004M8 |
| GH-LST | -5.55** | 2009M12 | -5.60** | -47.3 | 2010M2 |
| GH-RS | -3.91 | 2005M2 | -4.00 | -27.0 | 2008M8 |
| GH-RST | -5.46 | 2009M12 | -5.56 | -49.8 | 2009M4 |
| **DJUSFN** | | | | | |
| GH-LS | -2.95 | 2009M4 | -3.03 | -13.8 | 2009M3 |
| GH-LST | -5.64** | 2008M5 | -5.53** | -48.4 | 2008M8 |
| GH-RS | -4.26 | 2008M8 | -4.37 | -32.3 | 2008M8 |
| GH-RST | -5.44 | 2009M12 | -5.42 | -44.0 | 2009M5 |
| **DJUSHC** | | | | | |
| GH-LS | -3.32 | 2013M6 | -3.39 | -18.2 | 2013M4 |
| GH-LST | -4.89 | 2010M1 | -5.05 | -39.9 | 2010M2 |
| GH-RS | -4.20 | 2010M11 | -4.21 | -27.7 | 2010M10 |
| GH-RST | -5.52 | 2009M12 | -5.81* | -52.5 | 2010M2 |
| **DJUSIN** | | | | | |
| GH-LS | -3.27 | 2005M2 | -3.57 | -21.1 | 2004M6 |
| GH-LST | -5.34** | 2010M1 | -6.14*** | -58.6** | 2008M7 |
| GH-RS | -4.00 | 2008M6 | -4.47 | -34.2 | 2008M7 |
| GH-RST | -5.72 | 2009M12 | -6.06** | -56.6 | 2009M4 |
| **DJUSEN** | | | | | |
| GH-LS | -3.73 | 2005M1 | -3.87 | -24.8 | 2005M3 |
| GH-LST | -4.68 | 2008M2 | -4.94 | -38.8 | 2008M3 |
| GH-RS | -5.38* | 2006M12 | -5.15 | -42.0 | 2007M1 |
| GH-RST | -5.64 | 2009M10 | -5.71 | -53.9 | 2009M10 |
| **DJUSTC** | | | | | |
| GH-LS | -3.51 | 2005M2 | -4.15 | -27.1 | 2004M6 |
| GH-LST | -5.67** | 2008M4 | -5.58** | -49.2* | 2008M7 |
| GH-RS | -4.34 | 2005M2 | -4.57 | -34.6 | 2008M7 |
| GH-RST | -5.74 | 2009M12 | -6.09** | -57.8 | 2009M4 |
| **DJUSTL** | | | | | |
| GH-LS | -2.62 | 2012M10 | -3.31 | -18.7 | 2004M12 |
| GH-LST | -4.41 | 2008M4 | -5.53** | -52.7* | 2008M7 |
| GH-RS | -3.69 | 2008M4 | -4.78 | -38.7 | 2008M7 |
| GH-RST | -5.76* | 2009M11 | -7.01*** | -72.4** | 2009M10 |
| **DJUSUT** | | | | | |
| GH-LS | -3.01 | 2005M2 | -3.20 | -18.6 | 2004M12 |
| GH-LST | -6.03*** | 2008M5 | -6.22*** | -57.2** | 2008M7 |
| GH-RS | -4.29 | 2008M6 | -4.38 | -35.5 | 2008M7 |
| GH-RST | -6.64*** | 2008M5 | -6.81*** | -66.3* | 2009M10 |

*Notes: (i) \*, \*\* and \*\*\* implies significance at 10%, 5% and 1% levels respectively and rejection of no cointegration null hypothesis; (ii) the natural log of stock indexes is considered.*

As highlighted in Table 2, a cointegration relationship can be noticed for eight out of ten indexes. In the case of DJUSHC and DJUSEN no long-run relationship is documented.

Consequently, the health care sector and the energy stock prices are not influenced neither by the level of inflation,nor by its uncertainty in the long-term. These results can be explained by the fact that health care industry is not sensitive to macroeconomic fundamentals given its particularities, while in the case of energy sectorthe lack of a long-run relationship can be explained by the fact that energy prices are not completely "deregulated" in the U.S. In addition, an increase of oil and gas prices automatically leads to a rise of inflationary expectations. However, the effect of inflation is partially offset by higher nominal returns of energy stock prices.

For all other indexes, the cointegration relationship is documented by one or two tests (level shift with trend and regime shift with trend changes models). For three indexes (DJUSNC, DJUSTL, DJUSUT) both the second and the fourth model underlines the existence of a long-run relationship between stock prices, inflation and its uncertainty.

## *4.2. The long-run relationship*

Table 2 shows that in general one or two out of four models reject the null hypothesis of no cointegration for eight sector indexes. Therefore, in order to assess the long-run relationship, and to select the best model in the case of the three indexes where the cointegration is validated by the GH-LST and GH-RST tests (DJUSNC, DJUSTL, DJUSUT), we proceed to the estimation ofcointegration equations,using the Engle–Granger method and the OLS technique. The results are presented in Table 3 bellow.

The findings show that for those indexes for each the cointegration is documented base on the GH-LST model (DJUSCY, DJUSFN, DJUSIN, DJUSTC), both the inflation and the uncertainty negatively impact the stock prices in the long-run. Apparently in the case of basic materials and industrials sectors, even if the inflation coefficient sign is negative, it is not significant, which shows the dominance of the inflation uncertainty for the two indexes. Further, for those indexes where the cointegration is explained by two models (DJUSNC, DJUSTL and DJUSUT), there is rather the second model (GH-RST) which better explains the long-run relationship. In all cases the impact of inflation and its uncertainty is negative and very significant. However, we can notice that the negative impact is persistent until the moment of the structural break, which might be associated with the crisis period (the structural breaks are located between 2008M7 and 2010M2). Starting from this moment, the effects of inflation and its uncertainty practically disappear, results confirmed by a Wald test on coefficients.

Table 3. Cointegration equations between stock prices, inflation and uncertainty

|  | GH-LST | GH-RST |  | GH-LST | GH-RST |
|---|---|---|---|---|---|
| **DJUSBM** |  | *(Dum 2009M4)* | **DJUSIN** | *(Dum 2008M7)* |  |
| C |  | 4.963*** | C | 5.222*** |  |
| Dum × C |  | 0.200 | Dum × C | -0.522*** |  |
| Trend |  | 0.014*** | Trend | 0.010*** |  |
| Dum × Trend |  | -0.006*** | Dum × Trend |  |  |
| I |  | -0.013 | I | -0.000 |  |
| Dum × I |  | 0.026 | Dum × I |  |  |
| U |  | -0.383*** | U | -0.071*** |  |
| Dum × U |  | 0.307*** | Dum × U |  |  |
| **DJUSNC** | *(Dum 2008M7)* | *(Dum 2009M4)* | **DJUSTC** | *(Dum 2008M7)* |  |
| C | 5.437*** | 5.422*** | C | 5.935*** |  |
| Dum × C | -0.341*** | -0.754*** | Dum × C | -0.335*** |  |
| Trend | 0.008*** | 0.008*** | Trend | 0.009*** |  |
| Dum × Trend |  | 0.002*** | Dum × Trend |  |  |
| I | -0.040*** | -0.032*** | I | -0.020*** |  |
| Dum × I |  | 0.036*** | Dum × I |  |  |
| U | -0.235*** | -0.247*** | U | -0.092*** |  |
| Dum × U |  | 0.234*** | Dum × U |  |  |
| **DJUSCY** | *(Dum 2010M2)* |  | **DJUSTL** | *(Dum 2008M7)* | *(Dum 2009M10)* |
| C | 5.694*** |  | C | 4.641*** | 4.822*** |
| Dum × C | -0.477*** |  | Dum × C | -0.490*** | -0.465*** |
| Trend | 0.010*** |  | Trend | 0.006*** | 0.010*** |
| Dum × Trend |  |  | Dum × Trend |  | -0.004*** |
| I | -0.098*** |  | I | 0.007 | -0.062*** |
| Dum × I |  |  | Dum × I |  | 0.057*** |
| U | -0.394*** |  | U | -0.004 | -0.335*** |
| Dum × U |  |  | Dum × U |  | 0.213*** |
| **DJUSFN** | *(Dum 2008M8)* |  | **DJUSUT** | *(Dum 2008M7)* | *(Dum 2009M10)* |
| C | 6.243*** |  | C | 4.462*** | 4.635*** |
| Dum × C | -0.644*** |  | Dum × C | -0.509*** | -0.397*** |
| Trend | 0.003*** |  | Trend | 0.009*** | 0.012*** |
| Dum × Trend |  |  | Dum × Trend |  | -0.005*** |
| I | -0.067*** |  | I | 0.044*** | -0.021*** |
| Dum × I |  |  | Dum × I |  | 0.037*** |
| U | -0.265*** |  | U | 0.069*** | -0.274*** |
| Dum × U |  |  | Dum × U |  | 0.278*** |

*Notes: (i) \*, \*\* and \*\*\* implies significance at 10%, 5% and 1% levels respectively; (ii) standard errors in brackets; (iii) Dum means that the dummy variable is unity after this date and zero otherwise; (iv) the break is considered based on the $Z_t$ test.*

All in all, we conclude that our results point in general against the Fisher effect, and indirectly in the favor of the Friedman − Ball and Cukierman − Meltzer hypotheses, which sustain that the level of inflation and its uncertainty move in the same direction. However, after the crisis, in particular for the consumer goods and utilities indexes, the impact of inflation and its uncertainty became null, showing that the markets are less sensitive to the macroeconomic fundamentals. In the case of the utilities index these effect is reinforced by the fact that the demand is inelastic regarding the price movements.

After estimating and interpreting the cointegration relationships, we proceed to the estimation of the short-run relationship between stock prices, inflation and inflation uncertainty in the U.S., using an Error Correction Model (ECM).

*4.3. The short-run relationship*

To estimate the short-run dynamic equations we obtain the residuals from the representatives models (the GD-LST model for DJUSCY, DJUSFN, DJUSIN, DJUSTC and the GD-RST model for DJUSBM, DJUSNC, DGUSTL, DJUSUT). After obtaining residual series (called ECM), we can estimate the error correction models for the relationship between stock prices, inflation and inflation uncertainty (Table 4). For the two indexes for which there is no cointegration (DJUSHC and DJUSEN), the short-run relationship is computed based on a first differenced VAR model.

Table 4. ECM model and first differenced VAR estimation for stock prices, inflation and uncertainty

| ECM | DJUSBM | DJUSNC | DJUSCY | DJUSFN | DJUSIN | DJUSTC | DJUSTL | DJUSUT |
|---|---|---|---|---|---|---|---|---|
| C | 0.005 | 0.006** | 0.007** | 0.001 | 0.006* | 0.008* | 0.003 | 0.005* |
| $ECM_{t-1}$ | -0.194*** | -0.219*** | -0.107*** | -0.059** | -0.277*** | -0.297*** | -0.212*** | -0.194*** |
| $\Delta I$ | 0.013 | -0.005 | -0.015* | -0.016 | -0.007 | -0.001 | -0.015 | -0.000 |
| $\Delta U$ | -0.158*** | -0.104*** | -0.119*** | -0.230*** | -0.169*** | -0.167*** | -0.082* | -0.072** |
| $R^2$ | 0.137 | 0.139 | 0.114 | 0.106 | 0.259 | 0.206 | 0.108 | 0.144 |

| VAR | | $\Delta$DJUSHC | | | $\Delta$DJUSEN |
|---|---|---|---|---|---|
| | C | 0.008*** | | C | 0.006 |
| | $\Delta DJUSHC_{t-1}$ | 0.030 | | $\Delta DJUSEN_{t-1}$ | -0.045 |
| | $\Delta DJUSHC_{t-2}$ | -0.144 | | $\Delta DJUSEN_{t-2}$ | 0.126 |
| | $\Delta I_{t-1}$ | -0.000 | | $\Delta I_{t-1}$ | -0.024** |
| | $\Delta I_{t-2}$ | -0.001 | | $\Delta I_{t-2}$ | -0.016 |
| | $\Delta U_{t-1}$ | -0.162* | | $\Delta U_{t-1}$ | -0.159 |
| | $\Delta U_{t-2}$ | 0.110 | | $\Delta U_{t-2}$ | 0.002 |
| | $R^2$ | 0.052 | | $R^2$ | 0.085 |

*Notes: (i) *, ** and *** implies significance at 10%, 5% and 1% levels respectively; (ii) standard errors in brackets; (iii) $ECM_{t-1}$ is the lagged error correction term; (iv) All the VAR models retained into the analysis embody two lags and are stable.*

The ECM coefficient is negative and significant in all cases, which prove the existence of a long-run relationship between variables. However, the negative impact of inflation and its uncertainty on stock prices is less evident in the short-run. On the one hand the uncertainty determines a week negative impact, although significant. On the other hand, the inflation has no significant influence in the short-run for any of the eight indexes where a cointegration relationship exists. These results show that the macroeconomic fundamentals guide only the institutional, long-term investors, while the speculative traders being guided by the present economic context and by the uncertainty which characterize it. The first differenced VAR model computed for the DJUSHC and DJUSEN indexes shows no significant influence of inflation or its uncertainty in the short-run.

We proceed further to testing the stability of the analyzed relationship. In this respect we use a CUSUM test for the coefficient stability (Figure 1) and a CUSUM SQUARES test for the residuals stability (see Figure A1 – Appendix). The results show that all coefficients are stable in the case of ECM models, and a slightinstability in the residuals is recorded around the global crisis.

Figure 1. CUSUM test for the short-run relationship between stock prices, inflation and uncertainty

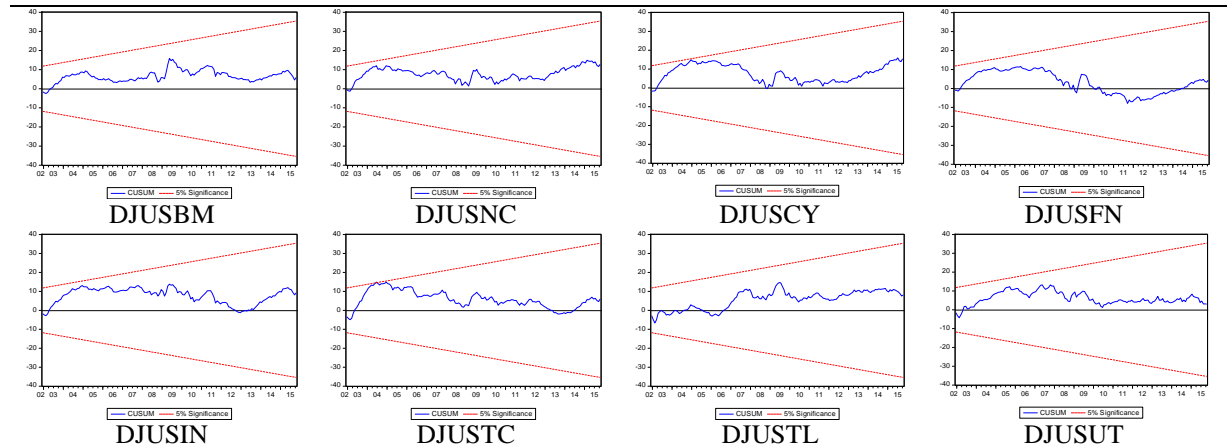

*Note: The straight lines represent critical bounds at 5% significance level.*

On the whole, we document that in the long-run both the inflation and its uncertainty negatively impact the stock prices, while in the short-run the stock pricesare influenced in particular by the inflation uncertainty. For two indexes, namely the DJUSHC and DJUSEN no significant influence was documented, while for three indexes (DJUSNC, DJUSTL and DJUSUT) a regime change is recorded. Therefore, for the consumer goods, telecommunication and utilities sectors the inflation and its uncertainty negatively influence the stock prices before the structural break, while a null effect is recorded after the crisis setup. However, these findings can be influenced by the phases of business cycle, fact which requires additional investigations.

## 5. Robustness check

For checking the robustness of our findings in the second step we consider the role of the business cycle. Thus, for the cointegration tests the general equation became:

$$DJ = c + \alpha_1 I + \alpha_2 U + \alpha_2 IP + \varepsilon_t \qquad (16)$$

where: $IP$ represent the industrial production growth rate.

As it can be noticed in Table 5, the results are practically unchanged. For two indexes (DJUSHC, DJUSEN) there is no long-run relationship, for three indexes (DJUSNC, DJUSTL, DJUSUT) both the GH-LST and GH-RST models provide evidence in the favor of cointegration relationship, while for the remaining sector indexes just one of the two models shows the existence of a long-run relationship.

Table 5. Gregory-Hansen cointegration tests for stock prices, inflation, uncertainty and output

| *Models* | ADF | Break | $Z_t$ | $Z_\alpha$ | Break |
|---|---|---|---|---|---|
| **DJUSBM** | | | | | |
| *GH-LS* | -4.43 | 2006M4 | -4.17 | -26.0 | 2006M5 |
| *GH-LST* | -5.38* | 2009M1 | -5.18 | -43.0 | 2010M1 |
| *GH-RS* | -5.35 | 2007M4 | -5.29 | -45.5 | 2008M7 |
| *GH-RST* | -6.66** | 2009M12 | -6.46** | -57.5 | 2009M1 |
| **DJUSNC** | | | | | |
| *GH-LS* | -3.94 | 2004M8 | -4.32 | -28.9 | 2004M6 |
| *GH-LST* | -6.34*** | 2009M12 | -6.39*** | -60.0** | 2010M1 |
| *GH-RS* | -4.54 | 2005M3 | -4.48 | -34.1 | 2008M7 |
| *GH-RST* | -6.18* | 2009M12 | -6.43** | -63.5 | 2009M4 |
| **DJUSCY** | | | | | |
| *GH-LS* | -3.92 | 2004M10 | -4.50 | -30.7 | 2004M8 |
| *GH-LST* | -5.59** | 2009M12 | -5.64** | -47.3 | 2010M2 |
| *GH-RS* | -4.53 | 2005M3 | -4.59 | -32.6 | 2011M1 |
| *GH-RST* | -5.45 | 2009M12 | -5.83 | -54.5 | 2009M4 |
| **DJUSFN** | | | | | |
| *GH-LS* | -2.99 | 2009M4 | -3.07 | -14.7 | 2009M3 |
| *GH-LST* | -5.87** | 2009M12 | -6.12*** | -57.5* | 2010M1 |
| *GH-RS* | -4.32 | 2008M9 | -4.90 | -39.5 | 2008M7 |
| *GH-RST* | -5.64 | 2009M12 | -5.85 | -53.4 | 2009M5 |
| **DJUSHC** | | | | | |
| *GH-LS* | -3.29 | 2013M6 | -3.89 | -23.8 | 2004M8 |
| *GH-LST* | -4.30 | 2008M6 | -5.07 | -39.9 | 2010M2 |
| *GH-RS* | -4.34 | 2012M5 | -4.51 | -33.4 | 2011M1 |
| *GH-RST* | -5.91 | 2008M9 | -6.28* | -61.9 | 2010M2 |
| **DJUSIN** | | | | | |
| *GH-LS* | -3.57 | 2004M6 | -4.00 | -25.7 | 2004M7 |
| *GH-LST* | -6.28*** | 2009M12 | -6.59*** | -62.2** | 2009M12 |
| *GH-RS* | -4.64 | 2008M7 | -4.65 | -38.5 | 2008M7 |
| *GH-RST* | -5.90 | 2009M12 | -6.29** | -58.8 | 2009M4 |
| **DJUSEN** | | | | | |
| *GH-LS* | -3.67 | 2004M11 | -4.00 | -26.1 | 2005M3 |
| *GH-LST* | -4.78 | 2008M2 | -5.05 | -40.4 | 2008M3 |
| *GH-RS* | -5.99* | 2006M12 | -5.73 | -50.3 | 2007M1 |
| *GH-RST* | -6.16 | 2009M11 | -6.33** | -63.2 | 2009M11 |
| **DJUSTC** | | | | | |
| *GH-LS* | -3.97 | 2011M6 | -4.27 | -28.4 | 2004M6 |
| *GH-LST* | -5.70** | 2008M5 | -5.72** | -51.4 | 2010M1 |
| *GH-RS* | -4.68 | 2010M12 | -4.64 | -35.1 | 2008M7 |
| *GH-RST* | -5.83 | 2009M12 | -6.20* | -59.6 | 2009M4 |
| **DJUSTL** | | | | | |
| *GH-LS* | -2.75 | 2005M1 | -3.65 | -23.9 | 2004M7 |
| *GH-LST* | -4.45 | 2008M4 | -5.93** | -56.1* | 2009M11 |
| *GH-RS* | -4.39 | 2008M4 | -5.63 | -53.7 | 2008M7 |
| *GH-RST* | -5.72 | 2009M8 | -7.61*** | -84.7** | 2008M7 |
| **DJUSUT** | | | | | |
| *GH-LS* | -3.40 | 2004M8 | -3.70 | -23.2 | 2004M7 |
| *GH-LST* | -6.74*** | 2009M11 | -6.87*** | -71.7*** | 2009M11 |
| *GH-RS* | -5.32 | 2008M7 | -5.05 | -46.0 | 2008M7 |
| *GH-RST* | -6.69** | 2009M9 | -7.58*** | -79.1** | 2009M9 |

*Notes: (i) *, ** and *** implies significance at 10%, 5% and 1% levels respectively and rejection of no cointegration null hypothesis; (ii) the natural log of stock indexes is considered.*

Table 6 presents the cointegration equations results. Several findings can be reported. First, as a general result we notice that the influence of the inflation and its uncertainty on stock prices remains negative and significant, while the output has a positive effect. Therefore, these results confirm the Fama's (1981) argument, who sustains that stock prices are correlated with the business cycle, being negatively impacted by the level of inflation.

Table 6. Cointegration equations between stock prices, inflation, uncertainty and output

|  | GH-LST | GH-RST |  | GH-LST | GH-RST |
|---|---|---|---|---|---|
| **DJUSBM** |  | (Dum 2009M1) | **DJUSIN** | (Dum 2009M12) |  |
| C |  | 5.065*** | C | 5.251*** |  |
| Dum × C |  | 0.076 | Dum × C | -0.602*** |  |
| Trend |  | 0.015*** | Trend | 0.010*** |  |
| Dum × Trend |  | -0.010*** | Dum × Trend |  |  |
| I |  | -0.025* | I | -0.019*** |  |
| Dum × I |  | 0.030 | Dum × I |  |  |
| U |  | -0.593*** | U | -0.195*** |  |
| Dum × U |  | 0.522*** | Dum × U |  |  |
| IP |  | -0.016** | IP | 0.020*** |  |
| Dum × IP |  | 0.025** | Dum × IP |  |  |
| **DJUSNC** | (Dum 2010M1) | (Dum 2009M5) | **DJUSTC** | (Dum 2010M1) |  |
| C | 5.410*** | 5.347*** | C | 5.953*** |  |
| Dum × C | -0.357*** | -0.667*** | Dum × C | -0.373*** |  |
| Trend | 0.008*** | 0.007*** | Trend | 0.009*** |  |
| Dum × Trend |  | 0.002*** | Dum × Trend |  |  |
| I | -0.039*** | -0.023*** | I | -0.033*** |  |
| Dum × I |  | 0.027** | Dum × I |  |  |
| U | -0.210*** | -0.136*** | U | -0.162*** |  |
| Dum × U |  | 0.119*** | Dum × U |  |  |
| IP | 0.004* | 0.013*** | IP | 0.014*** |  |
| Dum × IP |  | -0.015** | Dum × IP |  |  |
| **DJUSCY** | (Dum 2010M2) |  | **DJUSTL** | (Dum 2009M11) | (Dum 2008M7) |
| C | 5.696*** |  | C | 4.686*** | 4.797*** |
| Dum × C | -0.476*** |  | Dum × C | -0.535*** | -0.267** |
| Trend | 0.010*** |  | Trend | 0.006*** | 0.011*** |
| Dum × Trend |  |  | Dum × Trend |  | -0.006*** |
| I | -0.098*** |  | I | -0.007 | -0.080*** |
| Dum × I |  |  | Dum × I |  | 0.069*** |
| U | -0.396*** |  | U | -0.143*** | -0.271*** |
| Dum × U |  |  | Dum × U |  | 0.131*** |
| IP | -0.000 |  | IP | 0.012*** | 0.012** |
| Dum × IP |  |  | Dum × IP |  | -0.027*** |
| **DJUSFN** | (Dum 2010M1) |  | **DJUSUT** | (Dum 2009M11) | (Dum 2009M9) |
| C | 6.114*** |  | C | 4.443*** | 4.544*** |
| Dum × C | -1.069*** |  | Dum × C | -0.572*** | -0.338*** |
| Trend | 0.007*** |  | Trend | 0.009*** | 0.012*** |
| Dum × Trend |  |  | Dum × Trend |  | -0.005*** |
| I | -0.076*** |  | I | 0.034*** | -0.014* |
| Dum × I |  |  | Dum × I |  | 0.033** |
| U | -0.334*** |  | U | -0.006 | -0.156*** |
| Dum × U |  |  | Dum × U |  | 0.173*** |
| IP | 0.033*** |  | IP | 0.024*** | 0.018*** |
| Dum × IP |  |  | Dum × IP |  | -0.016*** |

*Notes: (i) *, ** and *** implies significance at 10%, 5% and 1% levels respectively; (ii) standard errors in brackets; (iii) Dum means that the dummy variable is unity after this date and zero otherwise; (iv) the break is considered based on the $Z_t$ test.*

Second, the role of business cycle in explaining the stock prices is important for the financial sector, as well as for the telecommunication and utilities. If these results are not surprising for the first two mentioned sectors, in the case of utilities the result is intriguing given that the latter sector is *a priori* assumed to be the less cyclical, given the long maturities of contracts concluded in this sector. Third, for those indexes where the model with regime change explains the long-run relationship, we notice that the influence of inflation or production is no longer significant after the crisis. Forth, for the two important sectors in terms of market capitalization in the U.S., namely the health care and energy sector, no cointegration is documented in this case also.

Table 7 addresses the short-run relationship results and presents similar findings to those reported in Table 4. Considering the role of business cycles does not bring any changes for the short-run relationship. While the uncertainty has a negative and significant influence on stock prices, the role of inflation is less evident, except for DJUSCY and DJUSTL (the inflation coefficient for DJUSCY was also significant in Table 4, but only at 10% significance level). In addition, given the results of the cointegration equations, as expected, the influence of business cycle is not significant in the short-run. Further, the VAR in difference model estimated for DJUSHC and DJUSEN shows no significant impact of inflation, uncertainty or output, which proves once again the robustness of our findings.

Table 7. ECM model and first differenced VAR estimation for stock prices, inflation, uncertainty and output

| ECM | DJUSBM | DJUSNC | DJUSCY | DJUSFN | DJUSIN | DJUSTC | DJUSTL | DJUSUT |
|---|---|---|---|---|---|---|---|---|
| C | 0.005 | 0.006** | 0.007** | 0.001 | 0.006 | 0.007* | 0.003 | 0.005 |
| $ECM_{t-1}$ | -0.365*** | -0.258*** | -0.103*** | -0.031 | -0.132*** | -0.199*** | -0.410*** | -0.232*** |
| $\Delta I$ | 0.016 | -0.006 | -0.017** | -0.015 | -0.009 | 0.000 | -0.017** | -0.001 |
| $\Delta U$ | -0.229*** | -0.122*** | -0.146*** | -0.198*** | -0.180*** | -0.143** | -0.105** | -0.076** |
| $\Delta IP$ | -0.005 | -0.003 | -0.004 | -0.003 | -0.003 | -0.003 | -0.004 | -0.001 |
| $R^2$ | 0.279 | 0.184 | 0.123 | 0.089 | 0.132 | 0.127 | 0.234 | 0.119 |

| VAR | | $\Delta$DJUSHC | | | $\Delta$DJUSEN |
|---|---|---|---|---|---|
| | C | 0.008*** | | C | 0.006 |
| | $\Delta DJUSHC_{t-1}$ | 0.028 | | $\Delta DJUSEN_{t-1}$ | -0.027 |
| | $\Delta DJUSHC_{t-2}$ | -0.137 | | $\Delta DJUSEN_{t-2}$ | 0.124 |
| | $\Delta I_{t-1}$ | -0.002 | | $\Delta I_{t-1}$ | -0.023* |
| | $\Delta I_{t-2}$ | -0.000 | | $\Delta I_{t-2}$ | -0.017 |
| | $\Delta U_{t-1}$ | -0.145* | | $\Delta U_{t-1}$ | -0.147 |
| | $\Delta U_{t-2}$ | 0.134 | | $\Delta U_{t-2}$ | 0.008 |
| | $\Delta IP_{t-1}$ | 0.002 | | $\Delta IP_{t-1}$ | 0.005 |
| | $\Delta IP_{t-2}$ | 0.003 | | $\Delta IP_{t-2}$ | -0.002 |
| | $R^2$ | 0.063 | | $R^2$ | 0.085 |

*Notes: (i) \*, \*\* and \*\*\* implies significance at 10%, 5% and 1% levels respectively; (ii) standard errors in brackets; (iii) $ECM_{t-1}$ is the lagged error correction term; (iv) All the VAR models retained into the analysis embody two lags and are stable.*

Finally, Figure 2 provides evidence in the favor of coefficients' stability for all the eight ECM models, while Figure 2A (Appendix) underlines in general the residuals' stability.

Figure 2. CUSUM test for the short-run relationship between stock prices, inflation, uncertainty and output

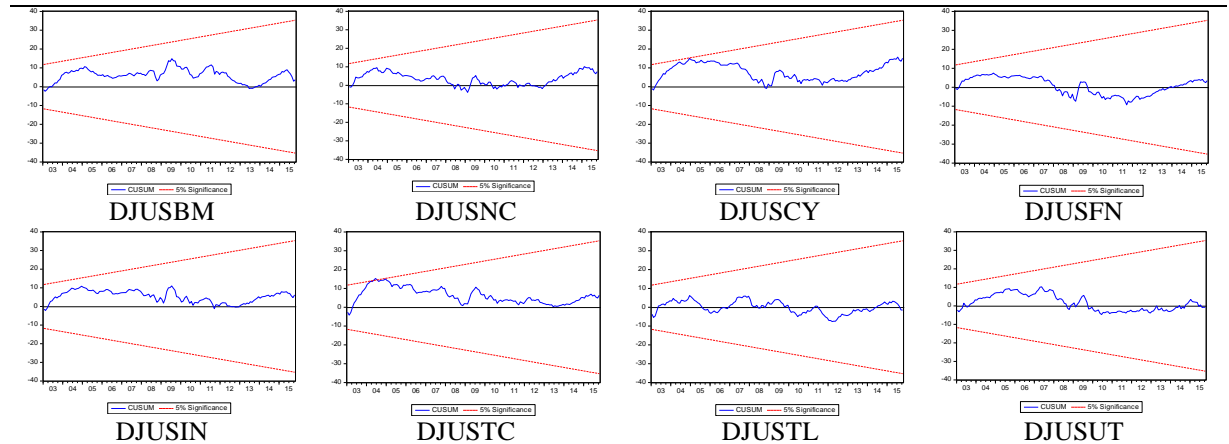

*Note: The straight lines represent critical bounds at 5% significance level.*

We conclude that, in the long-run there is a clear and robust relationship between our variables and in general the inflation and its uncertainty negatively impact the stock prices, results in agreement with the findings reported by Cochran and Defina (1993), Alexakis et al. (1996) and Karagianni and Kyrtsou (2011), and opposed to those advanced by Boudoukh and Richardson (1993) early on. The consideration of the output in our cointegration tests and equations confirms these results and the relatively strong negative impact of inflation uncertainty on stock prices (opposed to the findings reported by Azar, 2013). However, the models with regime shift shows that the effect of inflation and its uncertainty became insignificant in the long-run. In the case of the short-run equations the empirical evidence shows that the inflation uncertainty is negatively impacting the stock prices, while the inflation or the output has no influence. For two out of ten sector indexes (health care and energy), no long-run or short-run relationship is documented.

## 6. Conclusions

In this paper, we use the Gregory-Hansen cointegration tests to estimate the long-run relationship between stock prices, inflation level and inflation uncertainty in the U.S. The stock prices are estimated based on tenDJ sector indexes, during the period from 2002M7 to 2015M10. The inflation uncertainty is associated with the inflation gap and is computed based on the Stock and Watson's (2007) model with a time-varying inflation trend.

Our resultsprovide evidence for a long-run relationship between stock prices, inflation and its uncertainty, in the case of eight out of ten indexes. In particular, the cointegration relationship is shown by two Gregory-Hansen tests for DJUSNC, DJUSTL and DJUSUT (level shift with trend and regime shift with trend), and by one test for the remaining five indexes. The cointegration equations underline a negative impact of the inflation and its uncertainty on stock prices, meaning that they move together, indirectly supporting the Friedman – Ball and Cukierman – Meltzer hypotheses. However, the regime shift models highlight the lack of a significant impact of inflation and uncertainty on stock prices after the crisis setup. These results remain extremely robust after checking the impact of the business cycle.

The general long-term findings are then opposed to the Fisher's effect, but in agreement with recent findings documented by the empirical literature, and with the arguments advanced by Fama (1981), underlining the role of business cycles and the correlation existing between the real economy and equity markets. However, in the short-run the influence of the inflation became insignificant, while the uncertainty remains the only variable explaining the stock prices. The robustness analysis confirms these findings and underlines the fact that the production has no short-run influence on stock prices. Therefore, the fact that for several particular sectors (i.e. health care and energy) no relationship is documented shows that the results of previous studies, considering the general composite stock market indexes, should be considered with caution.

**References**

Alexakis, P., Apergis, N., Xanthakis, E., 1996. Inflation volatility and stock prices: Evidence from ARCH effects. International Advances in Economic Research, 2, 101–111.

Ammer, J., 1994. Inflation, inflation risk, and stock returns. FED International Finance Discussion Papers, no.464.

Azar, S.A., 2013. The Spurious Relation between Inflation Uncertainty and Stock Returns: Evidence from the U.S. Review of Economics & Finance, 3, 99–109.

Ball, L., 1992. Why does high inflation raise inflation uncertainty. Journal of Monetary Economics, 29, 371–388.

Bodie, Z., 1976. Common stocks as a hedge against inflation. The Journal of Finance, 31, 459–470.

Boudoukh, J., Richardson, M., 1993. Stock returns and inflation: a long-horizon perspective. American Economic Review, 83, 1346–1355.

Boyd, J.H., Levine, R., Smith, B.D., 2001. The impact of inflation on financial sector performance. Journal of Monetary Economics, 47, 221–248.

Campbell, J.Y. 1991. A variance decomposition for stock returns. Economic Journal, 101, 157–179.

Cochran, S.J., Defina, R.H., 1993. Inflation's negative effects on real stock prices: new evidence and a test for a proxy effect hypothesis. Applied Economics, 25, 263–274.

Cogley, T., Primiceri, G. and Sargent, T., 2010. Inflation-gap persistence in the U.S. American Economic Journal: Macroeconomics, 2, 43–69.

Cukierman, A., Meltzer, A., 1986. A theory of ambiguity, credibility, and inflation under discretion and asymmetric information.Econometrica, 54, 1099–1128.

Engsted, T., Tanggaard, C., 2002.The relation between asset returns and inflation at short and long horizons. Journal of International Financial Markets, Institutions and Money, 12, 101–118.

Fama, E.F., 1981. Stock returns, real activity, inflation, and money. American Economic Review, 71, 545–565.

Fama, E.F., Schwert, G.W., 1977. Asset returns and inflation. Journal of Financial Economics, 5, 115–146.

Feldstein, M., 1980.Inflation and stock market. American Economic Review, 70, 839–847.

Fisher, I., 1930. The Theory of Interest. New York: Macmillan.

Friedman, M., 1977. Nobel lecture: inflation and unemployment. Journal of Political Economy, 85, 451–472.

Geske, R., Roll, R., 1983. The fiscal and monetary linkage between stock returns and inflation. The Journal of Finance, 38, 1–33.

Gregory, A.W., Hansen, B.E., 1996a. Residual-based tests for cointegration in models with regime shifts. Journal of Econometrics, 70, 99–126.

Gregory, A.W., Hansen, B.E., 1996b. Tests for cointegration in models with regime and trend shifts. Oxford Bulletin of Economics and Statistics, 58, 555–559.

Hess, P., Lee, B-S., 1999. Stock returns and inflation with supply and demand disturbances. Review of Financial Studies, 12, 1203–1218.

Holland, S., 1995. Inflation and uncertainty: tests for temporal ordering. Journal of Money, Credit and Banking, 27, 827–837.

Hu, X., Willett, T., 2000.The variability of inflation and real stock returns.Applied Financial Economics, 10, 655–665.

Jaffe, J.F., Mandelker, G., 1976. The 'Fisher Effect' for risky assets: an empirical investigation. The Journal of Finance, 31, 447–458.

Karagianni, S., Kyrtsou, C., 2011.Analysing the Dynamics between U.S. Inflation and Dow Jones Index Using Non-Linear Methods. Studies in Nonlinear Dynamics & Econometrics, 15, Article 4.

Kaul G (1987) Stock returns and inflation: the role of the monetary sector. Journal of Financial Economics, 18, 253–276.

Kim, J.-R., 2003. The stock return-inflation puzzle and the asymmetric causality in stock returns, inflation and real activity. Economics Letters, 80, 155–160.

Kim, S., In, F., 2005. The relationship between stock returns and inflation: new evidence from wavelet analysis. Journal of Empirical Finance, 12, 435–444.

Kolluri, B.R., 1982. Anticipated price changes, inflation uncertainty, and capital stock returns. Journal of Financial Research, 5, 135–149.

Kolluri, B.R., Wahab, M., 2008. Stock returns and expected inflation: evidence from an asymmetric test specification. Review of Quantitative Finance and Accounting, 30, 371–395.

Levi, M.D., Makin, J.H., 1979. Fisher, Phillips, Friedman and the Measured Impact of Inflation on Interest. The Journal of Finance, 34, 35–52.

Liu, D., Jansen, D., Li., Q., 2005. Regime switching in the dynamic relationship between stock returns and inflation.Applied Financial Economics Letters, 1, 273–277.

Maghyereh, A., 2006. The long-run relationship between stock returns and inflation in developing countries: further evidence from a nonparametric cointegration test. Applied Financial Economics Letters, 2, 265–273.

Modigliani, F., Cohn, R.A., 1979.Inflation, rational valuation and the market. Financial Analysts Journal, 35, 22–44.

Nelson, C.R., 1976. Inflation and rates of returns on common stocks. The Journal of Finance, 31, 471–483.

Park, K., Ratti, R., 2000. Real activity, inflation, stock returns, and monetary policy. Financial Review, 35, 59–78.

Pourgerami, A., Maskus, K., 1987. The effects of inflation on the predictability of price changes in Latin America: some estimates and policy implications. World Development, 15, 287–290.

Schotman, P.C., Schweitzer, M., 2000. Horizon sensitivity of the inflation hedge of stocks. Journal of Empirical Finance, 7, 301–305.

Sharpe, S.A., 2002. Reexamining stock valuation and inflation: the implications of analysts' earnings forecasts. The Review of Economics and Statistics, 84, 632–648.

Solnik, B., Solnik, V., 1997. A multi-country test of the Fisher model for stock returns. Journal of International Financial Markets, Institutions and Money, 7, 289–301.

Stock, J., Watson, M., 2007. Why has U.S. inflation become harder to forecast? Journal of Money, Credit and Banking, 39, 3–33.

Tiwari, A.K., Dar, A.B., Bhanja, N., Arouri, M., Teulon, F., 2015. Stock returns and inflation in Pakistan. Economic Modelling, 47, 23–31.

Tobin, J. (1958). Liquidity preferences as behavior toward risk. Review of Economic Studies, 26, 65-86.

**Appendix**

Figure 1A. CUSUM SQUARES test for the short-run relationship between stock prices, inflation and uncertainty

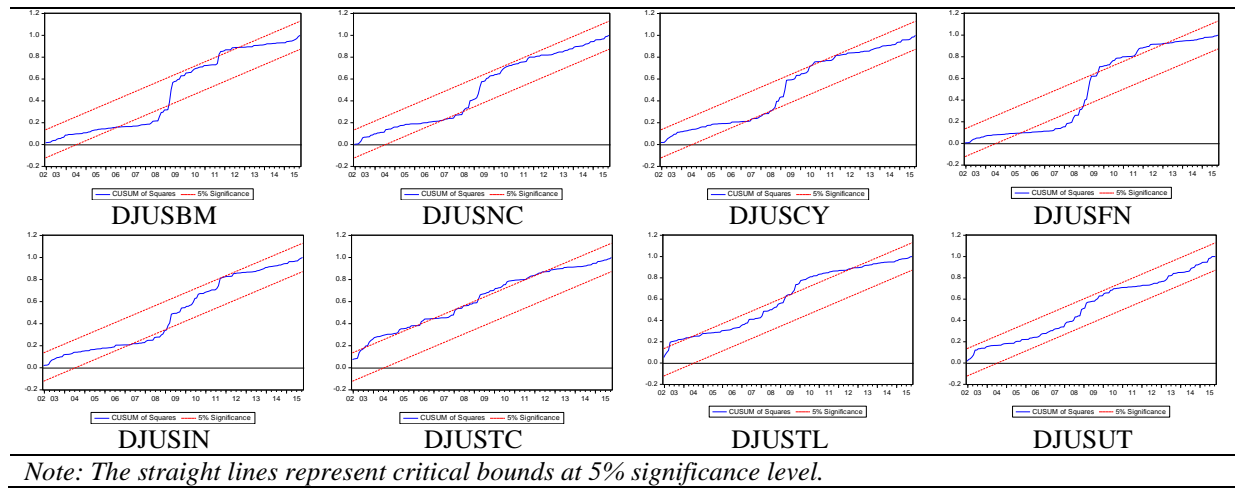

*Note: The straight lines represent critical bounds at 5% significance level.*

Figure 2A. CUSUM SQUARES test for the short-run relationship between stock prices, inflation, uncertainty and output

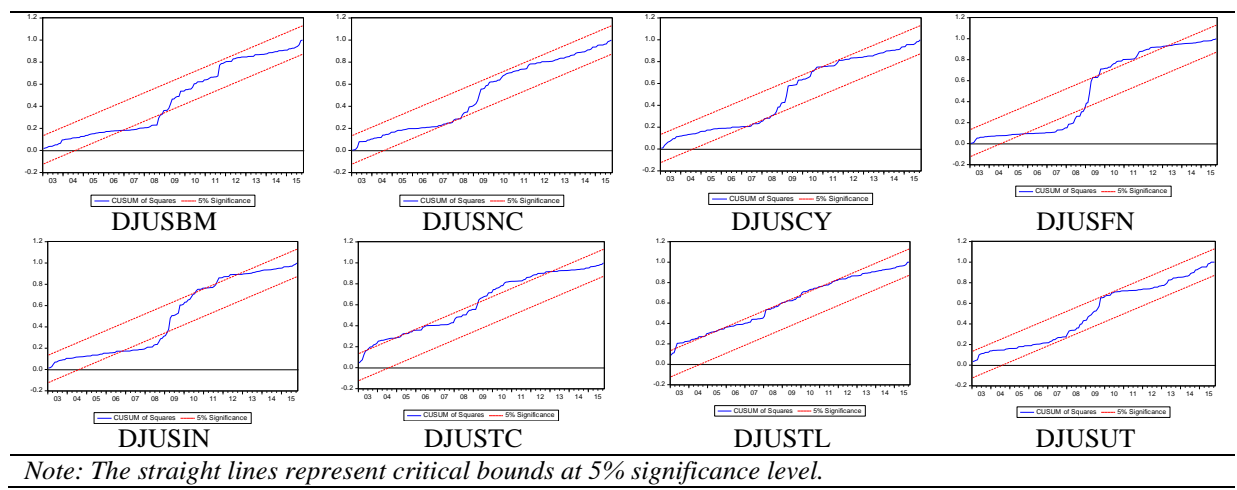

*Note: The straight lines represent critical bounds at 5% significance level.*